\documentclass{article}
\usepackage{fullpage}
\usepackage{amsfonts,amssymb}
\usepackage{amsthm}
\usepackage{tikz}
\usetikzlibrary{calc}
\usetikzlibrary{arrows}
\usepackage{tikz-3dplot}
\usepackage{color} 
\usepackage[fleqn]{amsmath}
\usepackage{amsthm}
\usepackage{amssymb}
\usepackage{amsfonts}
\usepackage{bbm}
\usepackage{epsfig}
\usepackage{times}
\usepackage{hyperref}
\usepackage{bm}
\usepackage{float} 
\usepackage{appendix} 
\usepackage{lscape}
\usepackage{cite}
\usepackage{mathrsfs}
\usepackage{appendix}
\usepackage{setspace}
\usepackage{mathtools}
\usepackage{mdframed}
\usepackage{authblk}
\theoremstyle{definition}

 \usepackage{hyperref}
  \usepackage{url}

\begin{document}

\title{ Are Entropy Bounds Epistemic? }

 \author{Emily Adlam  \thanks{The Rotman Institute of Philosophy, 1151 Richmond Street, London N6A5B7 \texttt{eadlam90@gmail.com} }}

\date{\today} 

\maketitle

\begin{abstract} 

Entropy bounds have played an important role in the development of holography as an approach to quantum gravity, so in this article we seek to gain a better understanding of the covariant entropy bound. We observe that there is a possible way of thinking about the  covariant entropy bound which would suggest that it encodes  an epistemic limitation rather than an objective count of the true number of degrees of freedom on a light-sheet; thus we distinguish between ontological and epistemic interpretations of the covariant bound. We consider the consequences that these interpretations might have for physics and we discuss what each approach has to say about gravitational phenomena.  Our aim  is not  to advocate for either the ontological or epistemic approach in particular, but rather to articulate both possibilities clearly and explore some arguments for and against them.

\end{abstract}

\section{Introduction}

Discussions about entropy bounds and holography  often focus on the strong entropy bound, which conjectures that the entropy associated with some spacelike volume is upper bounded by a quantity proportional to the surface area of the region. For example, holography is often described in terms of a map from space to the spatial boundaries of the world\cite{doi:10.1063/1.531249}. But there are known counterexamples to the strong entropy bound, including some which are believed to occur in our actual world; to fix these problems we must turn to the covariant entropy bound (also known as the Bousso bound), which tells us that the entropy associated with a null surface known as a `light-sheet' is upper bounded by a quantity proportional to the surface area of the spacelike region bounding the light-sheet. In Bousso's words, the covariant bound `\emph{evades these counterexamples because of the special properties of light-sheets. Thus, the notion of light-sheets (rather than spatial volumes, or light-cones lacking a nonexpansion condition), appears to be crucial.}' 

So it seems that if we are to understand the significance of entropy bounds and holography we should be focusing on the light-sheet construction used in the covariant bound. In this article we set out to understand the meaning of this construction. We observe that there is a possible interpretation of the  covariant entropy bound which would suggest that it encodes  an epistemic limitation rather than an objective count of the true number of degrees of freedom on a light-sheet; thus we distinguish between ontological and epistemic interpretations of the covariant bound. We discuss how these interpretations relate to Jacobson's derivation of the Einstein equations, and we consider the evidence for each interpretation. Our aim here is not  to advocate for either the ontological or epistemic approach in particular, but rather to articulate both possibilities clearly and explore some arguments for and against them. 

Note that this paper will largely be focused on entropy bounds in a semiclassical context - in a companion paper we will address the significance of the entropy bounds in the context of quantum mechanics and quantum gravity. Thus in this paper we will largely deal with entropy as a classical notion, deferring discussion of entanglement entropy until the companion paper.

\section{Background: Entropy Bounds \label{intro}}

Entropy bounds have their origins in an argument due to Bekenstein\cite{PhysRevD.7.2333} and Hawking\cite{cmp/1103899181} suggesting that if the  second law of thermodynamics is to be obeyed in the vicinity of black holes, we must generalize it to include an entropy associated with black holes equal to $\frac{A}{4   G \hbar}$, where $A$ is the surface area of the black hole; this value is known as the Bekenstein-Hawking entropy. Subsequently Bekenstein argued that if the generalized second law is to remain true when we lower systems into a black hole, it must be the case that the maximum entropy contained in a spherical system of radius $R$ and total gravitating energy $E$ is $S \leq \frac{2 \pi E R}{\hbar}$\cite{PhysRevD.23.287}; this  is known as the Bekenstein bound. Furthermore, for a weakly gravitating system  $E$ will always be significantly less than $\frac{2 \pi R}{4 G}$, so for weakly gravitating systems this bound reduces to the form  $S \leq \frac{A}{4 \hbar G}$, where $A$ is the area of a spherical surface fully containing the system.

Although this result was derived originally for the weak gravity regime, it  inspired a more general conjecture known as the `strong entropy bound' which says that the maximum entropy which can be contained in any region of space is upper bounded by $\frac{A}{4   G \hbar}$, where $A$ is the region of any surface bounding the region. For after all,  as originally pointed out by t'Hooft\cite{https://doi.org/10.48550/arxiv.gr-qc/9310026},if we could find a region with entropy greater than  $\frac{A}{4   G \hbar}$, we could add energy to it adiabatically without changing its entropy, thus transforming it into a black hole of area $A$ - but then this black hole would have entropy greater than the Bekenstein-Hawking entropy. So if the Bekenstein-Hawking entropy formula is correct, the entropy of a black hole of surface area $A$ must be the maximum possible entropy for any region of space with surface area $A$.  The strong bound  appears to be correct in many physically relevant situations\cite{PhysRevD.23.287}, and it is part of the inspiration for  a flourishing research field aiming to understand quantum gravity in terms of `holography'\cite{Maldacena_1999,https://doi.org/10.48550/arxiv.hep-th/9910146,Kolekar_2010}.

The strong entropy bound initially seems quite surprising. First of all, we may be surprised that there is any finite bound on the entropy inside a region of spacetime, since quantum field theory tells us that the number of degrees of freedom in any given region of space is infinite. There are ways we might make sense of this - for example, in quantum gravity, states above a certain energy can be expected to collapse to a black hole, and this mechanism might be expected to lead to a finite space of physically possible states\cite{Bao_2017}. Alternatively, several approaches to quantum gravity postulate that spacetime is discrete\cite{cc,rovelli_2004,https://doi.org/10.48550/arxiv.gr-qc/0505023,Sorkingeometry}, which would also explain why there should be a finite bound on the entropy of a spacetime region. However, the natural way of discretizing spacetime would seem to suggest that the degrees of freedom inside a region should scale with its volume, not the surface area of its surface,  so why does the entropy content of spacetime seem to be so much lower than we would naturally expect?  

In fact, the strong bound is not universally true: the heuristic arguments used to motivate it only work for sufficiently static scenarios, so it can be violated in situations like cosmic inflation and gravitational collapse.   These scenarios are rare, but nonetheless they are  important if one hopes to use the bound to learn something profound about fundamental physics, as many physicists apparently do - for a bound that is not universally valid cannot translate straightforwardly to a universal constraint on fundamental physics.  
  
  To solve these problems Bousso proposed a    `covariant entropy bound' \cite{Bousso_1999}: instead of bounding the information inside a volume enclosed by the relevant surface, it bounds the information on a `light-sheet' associated with the relevant surface. A light-sheet is  a set of null geodesics leaving the surface orthogonally such that the expansion of the set in the direction going away from the surface is zero or negative, i.e. the geodesics are remaining parallel or coming closer together as they get further from the surface. The light-sheet continues up until the geodesics intersect at a `caustic' (i.e. crossing-point) or encounter a singularity of spacetime. Bousso's bound then simply says that the entropy on the light-sheet associated with a surface of area $A$ is upper bounded by  $\frac{A}{4   G \hbar}$; and it currently seems likely that this bound is satisfied everywhere in our actual universe, at least at the semiclassical level\cite{Flanagan_2000}.

In the case of a closed sphere we can think of the null geodesics leaving the surface as the possible paths for rays of light leaving this surface;  they must  be emitted at the time $t$ and must be orthogonal to the surface of the sphere. In this case there are four possible directions in which the rays could be emitted - inside the sphere into the past, outside of the sphere into the past, inside the sphere into the future, or outside of the sphere into the future. Clearly the rays emitted outside  of the sphere are not converging, so these sets of rays have positive expansion and thus do not define light-sheets; but the rays emitted into the interior of the sphere are converging, so the sphere will have one interior light-sheet oriented into the past and another interior light-sheet oriented into the future. In the co-moving reference frame (assuming locally flat spacetime), the light emitted inwards will propagate towards the centre of the sphere and all of the rays of light will meet at the centre of the sphere at the same time. This meeting point is a `caustic' and thus the light-sheet terminates at this point. 

This example makes it easy to see why the covariant   entropy bound implies the strong entropy bound whenever the state inside the relevant surface is static. For in the co-moving reference frame, every spatial point inside the sphere is traversed exactly once by a ray belonging to the light-sheet, though different points will in general be traversed at different times. Thus if the state inside the sphere is static, the entropy traversed by these light rays is simply the same as the entropy of the state inside the sphere on any spacelike hypersurface, and thus the covariant bound implies the strong entropy bound in this  case.

\section{Does the Covariant Bound Count Degrees of Freedom? \label{count}} 

Entropy bounds have often been used to motivate claims about the number of degrees of freedom in some region of space. But care must be taken with these claims, because entropy does not straightforwardly correspond to an `amount of information' or `number of degrees of freedom.' For a start, these terms are quite vague - as Timpson stresses in ref \cite{10.1093/acprof:oso/9780199296460.001.0001}, there are a number of different things that the term `information content' could correspond to, and even more so in the context of quantum mechanics.  In addition, the validity of these claims seems to depend crucially on what kind of entropy one has in mind. The notion that  maximum entropy is a measure of a number of degrees of freedom seems to draw on an information-theoretic notion of entropy, i.e. the Shannon entropy,  which is given   by $\sum_x p(x) \log( p(x))$ where the sum is taken over all possible values of a variable $x$ and $p(x)$ gives the probability for some particular value. This is because if we maximize the Shannon entropy for all possible probability distributions over the possible states of a given physical system, the result can indeed be interpreted as counting the capacity for information storage in that system - for example,   given a system consisting of $n$ bits (i.e. $n$ independent systems with exactly two states each), the maximum Shannon entropy in units of bits is $n$. So talk of `degrees of freedom' in this context may be inspired by the idea of literally counting bits in simple systems, and indeed Bekinstein did suggest thinking about the black hole entropy in terms of Shannon entropy in his original paper on the subject\cite{PhysRevD.7.2333}

However, there are reasons to think that the Bekinstein-Hawking entropy should be understood primarily as a thermodynamic entropy, defined in terms of macroscopic thermodynamical quantities\cite{prunkl2019black,wallace2018case} - in particular, one of Bekinstein's initial arguments for the black hole entropy formula was that black holes must have this entropy in order to protect the second law of thermodynamics, and one of the main pieces  of (theoretical) evidence for the correctness of Bekinstein-Hawking formula is the fact that black holes can be shown  to emit Hawking radiation at exactly the temperature predicted by the first law of thermodynamics for a system whose entropy is the Bekinstein-Hawking entropy\cite{cmp/1103899181}. And since the entropy bounds were originally a kind of generalization of the Bekinstein-Hawking entropy, one might naturally think that they too should be understood first and foremost as thermodynamic entropies. Yet thermodynamical definitions of entropy in terms of macroscopic thermodynamical properties like mass, volume, pressure and temperature do  not necessarily seem to have anything to do with degrees of freedom. Of course, the thermodynamic entropy is usually  assumed to be reducible to the Gibbs and/or Boltzmann statistical-mechanical entropies (although even for paradigmatic thermodynamical systems the details of the reduction remain somewhat controversial\cite{cf4acd8e-334e-31c7-b917-af1b23023bf6,pittphilsci11019,doi:10.1142/9789811211720_0015}) and indeed there have been many proposals  for statistical-mechanical accounts of the Bekinstein-Hawking entropy within various possible approaches to quantum gravity\cite{https://doi.org/10.48550/arxiv.gr-qc/9705006, Strominger_1996, wallace2018case}. But since we don't yet have an empirically-verified theory of quantum gravity, there is still some doubt about the correct state space for a black hole, and thus it is still not certain that the black hole entropy can be thought of in a statistical-mechanical way. And likewise  for  the covariant bound: we don't yet know the complete state space for arbitrary regions of spacetime, since this state space may have to include quantum-gravitational degrees of freedom, so  perhaps we should not too quickly assume that the entropies featuring in entropy bounds can always be explicated as statistical-mechanical entropies\footnote{Of course, the entropy bounds might also be understood as relating to quantum-mechanical von Neumann entropies; we will discuss this in Part II of this paper.}. 

Furthermore, even if we decide the entropy appearing in the covariant bound \emph{can} be explicated as a statistical-mechanical entropy, there are still difficulties with identifying maximum statistical-mechanical entropies with degrees of freedom. For example, the Gibbs entropy is defined as $ - k_B \int_x p(x) \log( p(x))$ where $k_B$ is Boltzmann's constant, and $p(x)$ is a probability distribution over microstates $\{x\}$, which may be understood as representing an observer's knowledge about the true microstate $x$, or some kind of coarse-graining, or relative frequencies in an ensemble. This has a similar form to the Shannon formula, and indeed the maximum theoretically possible Gibbs entropy for a given system is  equal to the total number of possible microstates for that system, so there is a sense in which the maximum Gibbs entropy may be regarded as counting degrees of freedom - but note  that this maximum can be reached only with a probability distribution assigning equal probability to all possible microstates, which may or may not be a physically realistic probability distribution, depending on what we take the distribution to encode. Similarly,  the Boltzmann entropy is given by $k_B \ln(W)$ where $W$ is the total number of microstates compatible with the current macrostate of a system, so it is equal to the Gibbs entropy in the case where the probability distribution $p$ assigns equal probability to all the microstates compatible with the current macrostate (as for a system in thermal equilibrium using the canonical ensemble\cite{cf4acd8e-334e-31c7-b917-af1b23023bf6}), and is also equal to the Shannon entropy for a variable with $W$ equally probable values. Thus if there exists  a macrostate compatible with \emph{all} possible microstates,  its entropy will be the maximum possible Boltzmann entropy and that  entropy will indeed be something like a count of the total number of degrees of freedom - but note that  depending on how we identify macrostates, it may or may not be physically realistic to suppose that there is a macrostate which is compatible with all or nearly all possible microstates. So the maximum Gibbs and Boltzmann entropies are not guaranteed to yield the actual total number of microstates or `degrees of freedom' - this depends on the way in which we choose  the probability distributions or macrostates that we are maximizing over.

A further difficulty is presented by  the fact that thermodynamic entropy is arguably not an observer-independent quantity:  there is a long tradition, beginning with Maxwell\cite{maxwell1995maxwell}, of understanding thermodynamics as a science which involves designating a certain set of manipulable variables and then studying the responses of physical systems to manipulations of those variables.  For example, Myrvold\cite{Myrvold_2020}  argues for an an observer-dependent account of thermodynamics, arguing that entropy `\emph{must be relativized to the means we have available for  gathering information about and manipulating physical systems,}'\cite{pittphilsci8638}. The same can also be said of the statistical-mechanical definitions of entropy, since both the probability distribution $p(x)$ for the Gibbs entropy, and the choice of macrostates for the Boltzmann entropy, can also potentially be observer-dependent. And if we look more closely at the arguments for the strong entropy bound, it is striking that these arguments do indeed proceed by considering the processes that can be performed by agents who have access to only a certain set of `manipulable' variables (specifically, variables which can be accessed entirely outside of the black hole or relevant spatial region) - for example, Bekenstein's original argument for the Bekenstein-Hawking entropy involved  an assumption that entropy is `lost' once it enters the black hole and becomes inaccessible to external observers, and the argument for the strong entropy bound invoked a process in which an external  observer lowers a system into a black hole\cite{PhysRevD.23.287}.

However, anticipating this criticism, ref \cite{Flanagan_2000} makes the following argument: `\emph{there is an apparent tension between the fact that these statements are supposed to have the status of fundamental laws and the fact that entropy is a quantity whose definition is coarse-graining dependent. However, this tension is resolved by noting that the number of degrees of freedom should be an upper bound for the entropy S, irrespective of choice of coarse-graining.}' The idea here is that even though thermodynamical and statistical-mechanical entropies may be observer-relative, nonetheless if  we discover a universal bound on one of these entropies which applies to all possible observers, we have good reason to infer that this bound reflects the true number of degrees of freedom of the system, since no observer can assign to a system an entropy greater than its total number of degrees of freedom. Thus we will for now entertain the possibility that the covariant bound can indeed be regarded as providing a count of the total number of degrees of freedom in some region of spacetime. We will therefore be agnostic about exactly which of the four types of entropy discussed above is the one relevant to the covariant bound, since it seems reasonable to expect the total number of degrees of freedom of a given system to be an upper bound for all four types of entropy, so at least some potential explanations of the bound would seem to apply regardless of which type of entropy one has in mind.  

\subsection{Which region?}

Even if we accept that the covariant bound is counting degrees of freedom in some region of space, the question remains -  \emph{which} region?  Bousso himself\cite{Bousso_2002} infers from the covariant entropy bound  that the total number of degrees of freedom \emph{on a light-sheet} is proportional to the area of the surface with which the light-sheet is associated; we will refer to this as   \textbf{the light-sheet postulate}. On the other hand, others\cite{Smolin2001TheSA,   Jacobson_1999, https://doi.org/10.48550/arxiv.1710.00218, https://doi.org/10.48550/arxiv.gr-qc/9705006} advocate a weaker principle which says only  that the total number of degrees of freedom \emph{associated with a surface} is proportional to the area of that surface; we will refer to this as \textbf{the surface postulate}.  For example, Smolin gives a quantum formulation of the surface postulate that he calls the `weak entropy bound' which stipulates that for a system $\Sigma$ defined by the identification of a fixed boundary $\delta \Sigma = S$, and a Hilbert space $H_S$ defined as the the smallest faithful representation of the algebra of observables  measurable on the boundary only, then if the area $A(S)$ of the boundary is fixed, we must have $\log(dim(H_S)) \leq  \frac{A(S)}{4\hbar G}$\cite{Smolin2001TheSA}.

Smolin's argument for the surface postulate\cite{Smolin2001TheSA} is based on the fact that, as we have already noted, the entropies featuring in the derivation of the Bekenstein-Hawking entropy are thermodynamic ones; but rather than concluding on this basis that the covariant bound does not count degrees of freedom at all, he argues that since these thermodynamic arguments pertain specifically to exchanges of matter and  radiation between the black hole and external observers, the resulting bound must pertain to the degrees of freedom on the surface rather than the interior. Similarly, Sorkin notes that `\emph{the internal dynamics of a black hole ought to be irrelevant to its exhibited entropy because - almost by definition - the exterior is an autonomous system for whose behavior one should be able to account without ever referring to internal black hole degrees of freedom.}' On this basis Smolin\cite{Smolin2001TheSA}, and also Jacobson\cite{Jacobson_1999}, Rovelli\cite{https://doi.org/10.48550/arxiv.1710.00218} and Sorkin\cite{https://doi.org/10.48550/arxiv.gr-qc/9705006} all contend  that the Bekenstein-Hawking entropy describes only  the amount of information an observer external to the black hole can gain about its interior from measurements made outside the horizon, not the actual amount of information inside the black hole. Smolin  then concludes\cite{Smolin2001TheSA} that since the arguments for the strong entropy bound are derived from the Bekenstein-Hawking entropy bound, it follows that this bound also applies only to observables measurable on a boundary, not observables measurable in the interior. And since the covariant bound is intended as a generalization of the strong entropy bound, a similar argument would suggest that it too applies only to observables measurable on a boundary.  

But if  the surface postulate  is in fact the only inference about degrees of freedom justified by the covariant entropy bound, this would be a little disappointing, for the surface postulate doesn't seem to have much to do with `holography' in the usual sense. After all,  if for any reason we decide that surfaces in spacetime should be associated with a finite number of degrees of freedom, it would seem very natural to expect that   this finite number will be proportional to the area of the surface, so the surface postulate doesn't seem to give us much new information, other than possibly providing additional support for the idea that spacetime regions generally contain a finite number of degrees of freedom. Indeed, Smolin sees it as evidence for the hypothesis that spacetime is discretized:   `\emph{It is then difficult to escape the conclusion that the holographic principle, in its weak form, is telling us that nature is fundamentally discrete. The finiteness of the information available per unit area of a surface is to be taken simply as an indication that fundamentally, geometry must turn out to reduce to counting.}'\cite{Smolin2001TheSA}\footnote{We think Smolin's conclusion seems a little too strong here - a discrete spacetime could certainly explain why the number of degrees of freedom on a surface is finite, but there are  other possible explanations, such as the fact that states of too high energy will collapse to form a black hole, so it's unclear that we can immediately infer the existence of spacetime discreteness from finiteness.}
So if entropy bounds only tell us about the degrees of freedom on a surface,  they seem much less novel - as we have noted, the entropy bounds attracted so much interest precisely because they seemed to be pointing to something \emph{more} than mere finiteness or discreteness.

However,  there are reasons to think that the surface postulate does not exhaust the content of the entropy bounds.  For the surface postulate does not acknowledge the difference  between the original strong entropy bound and the covariant entropy  bound, and yet we seem to have reasonably good evidence that the covariant bound holds everywhere in our actual universe, at least in the semiclassical regime, while there are counterexamples indicating that the strong bound does not hold everywhere in our actual universe, even in the semiclassical regime.   So perhaps \emph{both} the surface postulate and the light-sheet postulate are correct?  Yet if they are both correct, we have some explaining to do,  because it surely cannot be a coincidence that the total number of degrees of freedom on a light-sheet are always related in this specific way to the total number of degrees of freedom on its bounding surface - it seems natural to think that these bounds must be connected in some way. We will now examine one interesting way in which they could be connected.

\subsection{Thought Experiment: Learning About Physics Inside a Region \label{thought}} 

In this section we will be concerned with issues relating to information and information transfer. So it should be emphasized that throughout this paper we will be focused on information as a technical, physical notion - we will concern ourselves with the ways in which physical systems exchange information and not, for example, the extent to which  information may be connected to language or meaning. In addition, we will follow  Timpson\cite{10.1093/acprof:oso/9780199296460.001.0001} in maintaining that information  `\emph{does not refer to a substance or to an entity'} - rather we understand information as an abstract noun which can be used to characterize various features of physical reality, including the ways in which  systems become correlated with each other, the inferences that can be made about one system from knowledge of the state of some other system, the efficiency with which certain kinds of communication systems   convey messages, and so on. So in this paper when we talk about information transfer or flux, we are not describing how some \emph{thing} travels: rather we are characterising physical processes in which physical systems move around and interact and thereby the  information stored in them becomes accessible at different locations in spacetime. In particular, in this article we are largely concerned with what Timpson refers to as \emph{accessible information} - i.e. information that can be accessed by a physically realistic observer performing physically possible measurements - and thus when we say that information has been transferred to some location, this entails  that the information can be accessed by observers at that location. On this understanding of `information transfer,' our best current physics makes it clear that information can be transferred between spacetime points only by local processes:  although we observe non-local effects in quantum physics, those non-local effects cannot be used to transfer accessible information (e.g. to send superluminal signals), and in fact all processes which do transfer information in this sense are mediated by physical signals travelling along continuous paths in spacetime. For example, in order to use quantum teleportation to transfer information via quantum entanglement to another observer, we must send a classical message to the other observer, and the amount of information that the other observer can ultimately obtain is no greater than the information content of the classical message\cite{PhysRevLett.70.1895}. 

So imagine that we have some information which is  initially located on the `inside’ of some boundary such as  a sphere, in a locally flat spacetime: now for any choice of reference frame we can identify a temporally extended 3D region corresponding to the location of the boundary (assuming it to be stationary in the chosen reference frame) and thus we can also identify a temporally extended 4D region corresponding to the interior of the boundary. And then, due to the locality of information transfer, the information originally on the inside can only get out of the temporally extended interior region (in that reference frame) by passing through the temporally extended boundary (in that reference frame).  In addition, the locality of information transfer seems to suggest  that this information transfer must involve a process in which the information in question can be regarded as occupying degrees of freedom  located on or near a spacelike slice through the temporally extended bounding surface, and therefore the number of degrees of freedom associated with each spacelike slice through the surface will constrain the rate at which information can move from the `interior' to the `exterior.' This will be true in any reference frame, though of course different reference frames will identify the interior and exterior regions differently, and then we can perhaps imagine taking limits as the temporal extension goes to zero, in order to arrive at an instantaneous rate of information transfer which does not depend on a choice of reference frame.  With this understanding of information transfer in mind, the surface postulate has a natural corollary: if the number of degrees of freedom associated with a given surface has an upper bound proportional to the area of the surface, then presumably  the rate at which information can pass through a surface also has an upper bound   proportional to the area of the surface.   

So now let us  consider a thought experiment.  Suppose you are presented with a large opaque sphere covered with $n$ LEDs. You are told that the interior of this sphere is not governed by the usual laws of physics - some completely new physics is going on inside it. The LEDs are linked to detectors on the interior, and they are reset at one second intervals, lighting up if their detector has received a `signal' from the inside during the most recent interval. You  have been assured that this is the only way in which information can emerge from the sphere. Your task is to use the information displayed by the LEDs to learn something about the physics on the inside. 

 Presumably, you will do something like the following. First, you will observe the sphere for a time, creating a record of the state of the LEDs every second. Then you will look at the series of states you have written down, attempting to discern some pattern. Hopefully you will eventually be able to write down some kind of `time evolution' law which relates past states to future states. Having done this, what more can you say about the physics inside the sphere?  

The simplest option is to postulate a one-to-one correspondence between states of the LEDs and states on the interior. Then the evolution law you arrived at for the LED states can be regarded as the evolution law for the states on the interior too; and clearly the states that you assign to the interior will be describable in exactly $n$ bits, which is to say they will have a Shannon entropy of no greater than $\log(n)$. You could perhaps do something more complex - for example, you could come up with a hypothesis about the nature of the signals which cause the LEDs to switch on or off, and thereby work backwards from the surface states to interior states. In this case there may no longer be a straightforward one-to-one mapping from the surface states to the interior states; however, it is still quite likely that the states you arrive at will be describable in $n$ bits or less, since you can get no more than $n$ bits of information out of the sphere at any given time to determine those states. 

We might make the thought experiment more complex by allowing you to `intervene,' as in a normal laboratory situation. This would involve a version of the scenario where the LEDs on the outside can both receive signals and also emit signals into the inside of the sphere, so you can prepare a `state' of the interior by setting the LEDs to some desired configuration, then let the system evolve and observe its response. But of course  you will only be able to prepare interior states up to a resolution of $n$ bits, since you only have $n$ degrees of freedom on which you can intervene, and then just as before you will only be able to learn $n$ bits about the succession of interior states that follows, so even with the ability to intervene it would not be surprising if you were to arrive at a description which assigns the interior only $n$ degrees of freedom. 

To be clear, we don't intend to suggest that you could \emph{never} be justified in assigning to the system a set of interior states which require more than $n$ bits to describe. After all, you are not just looking at a single observation but rather at a series of observations, and you might be able to find interesting correlations in the time-series data which you decide to  explain by postulating additional degrees of freedom inside the region\footnote{Thanks to an anonymous reviewer for pressing this point.}: for example, if you see non-Markovian features in the data, you might be inclined to think that there must be extra degrees of freedom inside the system storing additional information about the past. However, if you really have no idea at all about what kind of physics is going on inside the sphere, you will not know what kind of correlations to look for in the temporal data, so it could be quite hard to find them unless they take a particularly obvious form. In addition, even if you do see non-Markovian behaviour, it will always be open to you to suppose that the system just has  non-Markovian evolution laws rather than additional degrees of freedom, and arguably the former option is more ontologically economical. So although it is certainly possible for you to end up with a state space which requires more than $n$ bits to describe, nonetheless  it would not be in any way \emph{surprising} if you were to arrive at a description which assigns the interior only $n$ degrees of freedom, regardless of how many degrees of freedom there really are inside. Thus the point of this experiment is to show that  a bound of this kind  would be a natural outcome of the epistemic limitations you are subject to in this situation, even though it might be possible to do better than the bound with sufficient ingenuity.

We also note that if one is concerned by the possibility of gaining extra information from the temporal domain, it is possible to consider a generalization of this scenario using a different kind of surface, e.g. a temporally extended surface surrounding the entire sequence of observations that you extract from the LED sphere.  For example, consider Oeckl's general boundary formalism, which is motivated by considerations similar to those we have discussed here: as Oeckl writes,  for any arbitrary region in spacetime, `\emph{by locality, the interior of a region can interact with the rest of the universe only through the region’s boundary}'\cite{oeckl2014firstprinciples}. Oeckl therefore  argues that experiments can always be thought as being enclosed by a boundary such that the agent performing the experiment is outside of the boundary and can only manipulate and observe variables on the boundary; thus he constructs a reformulation of quantum mechanics designed to predict `transition amplitudes' from one part of the boundary to another.  Oeckl seems to suggest that one might expect something like an entropy bound to hold here: `\emph{connected boundaries of compact regions are the main focus of attention ...  in this sense the formulation is \textbf{holographic} i.e. the information about the interior of a region is encoded through the states on the boundary}'\cite{OECKL2003318}. So it is possible that the covariant entropy bound may ultimately be regarded as a special case of a temporally extended boundary formalism like Oeckl's, which would provide a more realistic model of our circumstances as temporally extended observers. However, in this article we will continue to focus on the covariant bound, as we think the instantaneous case is sufficient to explore a number of interesting ideas about the role of surfaces and the locality of information transfer.

\subsection{Light-sheets and accessible information}

The upshot of all this is that   your epistemic situation with regards to the sphere leads to a natural `entropy bound' -  it is natural to expect that the entropy you assign the interior may be upper bounded by the number of LEDs on the surface. Moreover, it would seem that the sphere example is a fairly close analogy for the situation described by the covariant entropy bound. 

In light of this, consider again the specific role played by the light-sheet construction in the covariant entropy bound. First note that a `null surface' is really the appropriate relativistically covariant way of describing the region to which we assign a state based on information received at a single time. For example, suppose I observe a football field. Although I may have the impression that I am observing the whole region simultaneously, that is not really true; each `snapshot' that I see of the field corresponds not to a spacelike hyperplane but rather  to something like a null surface, assuming that all the information reaches me at the speed of light. Now, of course I do not directly `see' the whole null surface, since it's possible for light to be blocked by opaque objects: what I see is something like an infinitesimal two-sphere encoding information from various different parts of the null surface. So the null surface can be thought of as containing all the information that would in principle be available to me in the best possible case, i.e. it is the locus of all the information I could possibly obtain from my observation if there were nothing blocking the light from reaching me.

What about light-sheets?  Well, they are simply a special kind of null surface. Specifically, a past-directed light-sheet is the locus of the information available to me if instead of collecting information at a single point, I collect it from a bounding surface. For example, if the  surface is a sphere and I collect all the information that arrives at the surface at a single instant of time in the co-moving reference frame, then assuming that the information all arrived at the surface at the speed of light and nothing blocks it, I get a `snapshot' of the interior of the sphere which corresponds to the associated light-sheet. So in a sense, the role of the light-sheet construction is simply to identify the appropriate null surface to which we should assign a state based on information collected at a spacelike surface; and as we have just seen in our thought-experiment,  it is natural under these circumstances to assign states from a state space which has no more than $n$ degrees of freedom, where $n$ is the number of bits which can pass through the  surface in one instant of time. And of course  if the surface postulate is correct, it follows that $n$ will be proportional to the area of the surface.  
 
One might object that this argument only applies to a closed surface, whereas the covariant entropy bound  applies also to open surfaces. But we can formulate a similar argument for an open surface. Suppose I am standing on one side of a screen of area $A$ and trying to gain information about a system on the other side of the screen: if the surface postulate is correct,  the rate at which information about the system can pass through the screen to me has an upper bound proportional to the area of the surface. Of course, in this case it's possible for information to go  around the screen in order to reach me, but if I am to learn anything from that information I'd have to know about the state of the region above around the screen so I can account for any further interactions the signals might undergo on the way. Thus we need to consider the process by which I learn the state of the surrounding regions as well, which requires us to extend the screen to cover information coming from those regions too, and so on, and therefore it's hard to see how I could do any better than learning a number of bits proportional to the area of the surface, unless I'm allowed to assume that I already know the state of some region around the screen. So the epistemic interpretation of the bound seems to work for open surfaces too: the light-sheet of such a surface still represents the locus of all the information which could pass through the screen in a given instant, which can be expected to yield an upper bound proportional to the area of the screen.

Our thought experiment is designed to highlight the  the special role of surfaces as interfaces through which observers access systems. This has been pointed out by various authors. For example, Oeckl\cite{OECKL2003318} has emphasized this feature of experimental design: `\emph{Consider for example a scattering experiment in high energy physics. A typical detector has roughly the form of a sphere with the scattering happening inside (e.g. a collision of incoming beams). The entries for particles and the individual detection devices are arranged on the surface}.'  Similarly Crane\cite{doi:10.1063/1.531240} observed that `\emph{if we divide the universe into ``system” and ``observer” the observer no longer measures the state of the system, but	only that part of it which impinges instantaneously on the observer,}' and thus proposed a system of observable algebras and Hilbert spaces, one associated with every possible splitting of the universe into system and observer. And  May\cite{2019JHEP...10..233M} has argued that holography can be expressed as the requirement that the asymptotic tasks which are possible using the bulk dynamics should coincide with tasks that are possible using the boundary; here `asymptotic tasks' are those which can be stated in terms of inputs and outputs located on the spacetime boundary. Put this way, it does not seem surprising  that our physics should have this feature: for after all we never really access a bulk directly, we only ever manipulate systems through a boundary, i.e.   the spatial boundary separating the the system from us, so it is to be expected that the physical theories we come up with describing possible tasks should be limited to the tasks which are possible using the boundary.

\section{Epistemic or Ontological \label{schema}}

The thought experiment of section \ref{thought} leaves us with a dilemma. For it suggests that, regardless of the number of degrees of freedom there really are on light-sheets, it is natural for observers to end up describing the physics on a light-sheet with a number of bits no greater than the amount of information which can pass through the associated surface - which according to the surface postulate will be proportional  to the area of that surface. So how can we ever know whether or not the covariant bound describes the true number of degrees of freedom on a light-sheet? It seems that anyone who accepts the surface postulate has good reason to think of the covariant bound  not as describing the true numbers of degrees of freedom on a light-sheet, but simply as an epistemic restriction on the number of degrees of freedom accessible to an external observer. We note that the possibility of the bound being epistemic in this sense is implicit in our discussion of the Gibbs and Boltzmann entropies in section \ref{count} - the   probability distributions or macrostates that we are maximizing over can be understood as being in some sense `epistemic' and we noted already that some ways of choosing the set of probability distributions or macrostates that we maximize over will lead to `maximum entropies' which are smaller than the total number of possible microstates or degrees of freedom.

So we now set out several different attitudes that one might take to the covariant bound, assuming that one accepts the surface  postulate:
 
\begin{enumerate} 

\item   \textbf{Ontological:} views which hold that the covariant entropy bound is a constraint on the number of real number of  ontological degrees of freedom on a light-sheet:

\begin{enumerate} 

\item \textbf{Pragmatic-ontological:} views which invoke some form of pragmatism or empiricism to say that since we can never have any information about degrees of freedom beyond the covariant entropy bound, we should assume there aren't any further degrees of freedom, but the bound doesn't have any deep meaning beyond that. 

\item \textbf{Positive-ontological:} views which postulate that the covariant entropy bound is a fundamental constraint or follows from aspects of the fundamental ontology, so it tells us something profound about the nature of our reality.

\end{enumerate}

\item  \textbf{Epistemic:} views which hold that the covariant entropy bound  is merely epistemic, and the real number of degrees of freedom on a light-sheet may be greater than the amount specified by the   bound: 

\begin{enumerate} 

\item \textbf{Pragmatic-epistemic:} views which postulate that there may exist degrees of freedom beyond those allowed by the covariant entropy bound, but we will probably never know anything about them and they will have no impact on anything we can observe, so we can get on with physics as if they don't exist at all.

\item \textbf{Positive-epistemic:} views which postulate that there may exist degrees of freedom beyond those allowed by the covariant entropy bound, and it may be important to take them into consideration, even though we can't directly observe them.

\end{enumerate}

\end{enumerate}

Now, there may appear to be a natural connection   between interpretations of the covariant bound   and the interpretation of the entropy appearing in the covariant bound. In particular, if this entropy is understood to be a thermodynamic or statistical-mechanical entropy which is regarded as being observer-relative in the sense described by Myrvold, then it seems natural to think of the covariant bound as a claim about the information which can be obtained by an observer with access to certain manipulable variables, so it is epistemic.  Moreover,  this identification may  appear to settle the question in favour of epistemic views, since as we have noted, the thermodynamic arguments for the Bekenstein bound seem to pertain to observer-relative thermodynamic entropies, and the covariant bound is a generalization of the Bekenstein bound. 

However, we would caution against making this identification in an overly simple way.  For proponents of an ontological view of the bound need not deny that the entropies employed in arguments for the Bekenstein-Hawking entropy and the strong entropy bound are observer-relative  entropies - it may be that \emph{in any particular case}, the entropy that we assign to a region of spacetime  is a measure of the information available to some specific (possibly hypothetical) observer.  But   the Bekenstein-Hawking entropy and the strong and  covariant bounds do not describe the entropy assigned in any particular case, rather they represent a universal upper bound on the entropies assigned by all possible observer in all possible cases. And the universality of this bound means that it cannot be understood as a measure of the information available to any one specific observer, so we still have the possibility of  interpreting it ontologically - for example, we have already mentioned the argument of Flanagan et al \cite{Flanagan_2000} that the universality of the bound indicates that it is a count of the true number of degrees of freedom. On the other hand, we have noted that a universal entropy bound could also be explained by   the fact that all realistic agents  are all subject to some kind of epistemic restriction which prevents them from gaining more than a certain number of bits of information about it. So really, nothing can be inferred one way or another from the observer-relative nature of the entropies used in the derivation of the bounds, because the nature of those entropies does not tell us the origin on the \emph{universal bound} on those entropies. Thus we would urge that, contrary to the arguments made by Smolin and others\cite{Smolin2001TheSA}, the thermodynamical nature of Bekenstein's reasoning does not settle the question of whether the covariant bound (or the strong entropy bound, or even the Bekenstein-Hawking entropy) is epistemic or ontological.

We can classify some of the views we have previously examined within this schema. Bousso clearly takes a  positive-ontological view, writing of the covariant entropy bound: `\emph{The bound’s simplicity, in addition to its generality, makes the case for its fundamental significance compelling}'\cite{Bousso_2002}. Other   proponents of holography in the conventional sense, such as Susskind\cite{doi:10.1063/1.531249} and t'Hooft\cite{https://doi.org/10.48550/arxiv.gr-qc/9310026}, also seem to take this kind of view. On the other hand Smolin\cite{Smolin2001TheSA}, Jacobson\cite{Jacobson_1999} and Sorkin\cite{https://doi.org/10.48550/arxiv.gr-qc/9705006} all appear to take an epistemic view. For example, these kinds of epistemic considerations are implicit in Smolin's criticisms of the original Bekenstein bound\cite{Smolin2001TheSA} - he emphasizes that the thermodynamic bounds on the entropy inside a black hole pertain specifically to the \emph{accessible} entropy, and we shouldn't assume that the  information accessible from outside a surface constitutes a bound on the information inside that surface, so it isn't clear that the arguments for Bekenstein's bound necessarily constrain the `real' degrees of freedom.

Meanwhile, it seems likely that the idea that entropy is `lost' when information falls into a black hole is partly grounded on a pragmatic view, since this would justify the practice of simply ignoring the possibility of degrees of freedom beyond the accessible ones. For example, Dougherty and Callender offer the following account of the reasons why physicists believe that entropy is lost when it crosses a black hole horizon: `\emph{it is held that the entropy vanishes when it passes behind the event horizon because we can’t gain access to it. The system itself doesn’t vanish; indeed, it had better not because its mass is needed to drive area increase. But for the ordinary entropy, when it crosses the event horizon it’s “out of sight, out of world,” or at least, out of physics.}'\cite{pittphilsci13195} So possibly some physicists are really agnostic about whether there is additional information in the black hole, but are willing to disregard  such information for  practical purposes, and presumably these physicists would also   be inclined to take a pragmatic-epistemic or pragmatic-ontological view to the covariant entropy bound.

 However, there is potential for the pragmatic approach to lead to confusion: if we simply ignore the existence of some degrees of freedom, we may get strange effects where information appears to mysteriously vanish. Indeed, it has been argued that this is exactly what is going on in the case of the black hole information paradox. As noted by Rovelli\cite{https://doi.org/10.48550/arxiv.1710.00218}, the Page argument for the information loss paradox `\emph{is based on the fact that if the number of black hole states is determined by the area, then there are no more available states to be entangled with the Hawking radiation when the black hole shrinks}' and thus `\emph{if there are more states available in a black hole than $e^{A/4}$, then the Page argument for the information  loss paradox fails.}'  Also, as  noted in section \ref{thought}, it is possible that even though we cannot directly access these extra degrees of freedom, we might be able to detect signatures of their existence by careful analysis of time-series data, for example via the detection of non-Markovian behaviour, so there may be important physical insights to be gained by taking into account the possibility of additional states. Thus although the pragmatic views have merits in certain contexts, we will henceforth focus on the positive views.
 
 \subsection{Positive-Epistemic Views \label{uncertainty}}

 It is common to understand the term ‘epistemic’ as connoting ‘not objective.’ For example, sometimes a distinction is made between ‘objective’ and ‘epistemic’ conceptions of entropy, as in ref\cite{pittphilsci13195}. But we emphasize that entropy bounds could be ‘epistemic’ while still being objective: for example, an entropy bound could arise from an objective description of the way in which certain kinds of physically embodied observers are limited in their ability to access certain kinds of information. That is to say, these bounds are `epistemic' not because they are merely features of the knowledge of one particular agent, but because they are objective descriptions of what it is possible for a certain kind of physically embodied agent to know, as a consequence of the nature of their physical embodiment and the means of enquiry available to them.  Moreover, we are using the term `epistemic' in quite a general way here: the putative epistemic interpretation of the covariant bound involves the suggestion that the bound pertains to the number of degrees of freedom which can be \emph{accessed} by an external observer, which is to say, it describes what the observer can know about the system, but it also describes the means available to the observer to act on, manipulate and extract work from the system. As argued by Robertson and Prunkl\cite{robertson_prunkl_2023}, resource-relativity of this kind is not `subjective,' even if it is to some degree anthropocentric.  So really ‘epistemic’ here should be understood as meaning something like ‘relational’ - if the covariant bound is epistemic, that simply means it doesn’t describe the degrees of freedom of an individual physical system, but rather says something about a relation between physical systems and observers, which themselves are a kind of physical system. These sorts of objective, relational epistemic facts are familiar from  thermodynamics - for example, Myrvold emphasizes that although  thermodynamic concepts like entropy are observer-relative, at the same time  ‘\emph{it would be misleading to call them subjective, as we are considering limitations on the physical means that are at the agents’ disposal.}'\cite{pittphilsci8638}

To understand what follows from an positive-epistemic view, let us disregard established physics   and ask   what  effects we might expect to observe if we did in fact live in a world where every region of space contains some   degrees of freedom which are inaccessible to any external observers. For a start, presumably we would observe phenomena  which would look indeterministic, even if the underlying processes are really deterministic: the inaccessible degrees of freedom would essentially act as `hidden variables,' since they would determine the outcomes of various interactions but we would never be able to observe them directly.  Thus it may be tempting to elide the hidden degrees of freedom with the `hidden variables' invoked in some interpretations of quantum mechanics, thus potentially providing a novel explanation for quantum indeterminism. In addition, we might expect to see some apparently non-Markovian behaviour as the inaccessible degrees of freedom could store extra information about the past, and since it has been argued that quantum mechanics itself has non-Markovian features\cite{2008PhRvA..77b2104M}, additional degrees of freedom could again play an explanatory role. 

 Second, perhaps we might get behaviour superficially similar to the uncertainty relations in quantum mechanics\cite{1927ZPhy...43..172H}. For suppose a particle is localised with position uncertainty $\delta x$ in each direction. Thus the particle is definitely to be found within a region of surface area $4 \pi \delta x^2$, and the surface postulate implies that the total amount of information we can get out of this region per unit time is proportional to  $4 \pi \delta x^2$. In particular, in order to find out the momentum of the particle we must get the information about its momentum at a given time out of the region, and thus the surface postulate implies that the smaller the region, the less information we can obtain about its momentum, or equivalently the greater the momentum uncertainty. So, roughly speaking, we'd expect that as the particle becomes better localised in spacetime our uncertainty about its momentum would increase, and vice versa.

 These examples show that an epistemic interpretation of the entropy bound may offer interesting new possibilities for understanding quantum mechanics. Indeed,   the bound on the amount of information which we can get out of a spacetime region looks a lot like the kind of `epistemic limitation' postulated by Spekkens, and as he shows in ref \cite{Spekkensepistemic}, the existence of epistemic limitations can explain many of the puzzling features of quantum mechanics. So  there may  indeed be value in thinking explicitly about the possibility of additional degrees of freedom beyond the covariant bound, rather than merely disregarding them on pragmatic grounds - even if we can't learn anything very concrete about these degrees of freedom,  simply acknowledging their existence can  provide novel explanations of various features of physics. Let us now consider whether the epistemic and ontological views of the covariant bound can shed any light on the matter.

 \subsection{Positive-Ontological Views \label{ontological}} 
 
The very possibility of an epistemic interpretation of the bound poses quite a serious problem for attempts to understand it in an ontological way, because proponents of such views will  be vulnerable to the criticism that they are simply mistaking an epistemic constraint for an ontological one. For example, the fact that four different methods of calculating the black hole entropy all give the Bekenstein-Hawking entropy formula\cite{https://doi.org/10.48550/arxiv.1710.02724} might initially be taken as evidence that the formula reflects  the true number of degrees of freedom in a black hole, but once we start thinking about possible epistemic restrictions, we might worry that the result is the same in all four cases not because it reflects the true, ontological number of degrees of freedom of the black hole but merely because all four approaches are based on physics developed by agents subject to a certain kind of universal epistemic restriction.

How can proponents of an ontological view resist this? Well, they could call into question some of the   ontological and metaphysical presuppositions which enter implicitly into the epistemic view. In particular, the argument that there could be more degrees of freedom inside a given region than observers are able to access is founded on the assumption that we inhabit a three-dimensional space (defined by a foliation of a four-dimensional spacetime) with independent autonomous degrees of freedom at each point of space.  But the thought experiment of section \ref{thought} makes it clear that this assumption far outstrips our evidence -  we don't actually receive data in three-dimensional chunks but rather on two-dimensional surfaces, so there is a significant mismatch between our  representation of the data  in 3D space and the amount of information we actually obtain about the contents of space. Thus rather than concluding that there exists a whole realm of invisible, inaccessible physics, perhaps we should adopt a different ontology which is a better match to the data. 

For example, one might imagine an ontology where information at distinct spacetime points  is not independent, so the total number of effective degrees of freedom scales instead with the surface area. As t'Hooft puts it:  `\emph{This suggests that physical degrees of freedom in three-space are not independent but, if considered at Planckian scale, they must be infinitely correlated}'\cite{https://doi.org/10.48550/arxiv.gr-qc/9310026}. Alternatively, one could imagine   an ontology composed entirely of two-dimensional surfaces, or something isomorphic to them. After all, since we always interact with bounding surfaces rather than the light-sheets associated with those surfaces, it seems that only the bounding surfaces are strictly necessary to account for our experiences, so we could potentially take it that   a bounding surface and its associated light-sheet are really  just equivalent representations of the same element of ontology. Indeed,   there are some indications from within General Relativity itself that this may be the right way to think about its ontology - for diffeomorphism invariance together with the equivalence principle prevent us from defining quantities like energy, momentum and angular momentum at individual spacetime points, and thus these quantities must instead be defined in terms of integrals  over two dimensional surfaces\cite{REGGE1974286}; so it seems quite natural to associate the theory not with an ontology of locally defined quantities at individual spacetime points, but rather with an ontology in which all meaningful quantities are associated with surfaces.

However, if we choose an ontology specifically to match the nature of the data available to observers like us, we run the risk of arriving at on ontology which works for the kinds of systems we observe but which depends on a split between systems and observers and hence does not allow us to accommodate observers or to explain the relation between systems and observers in physical terms. For example, it is easy enough to say that whenever an observer observes a region through a bounding surface, only the surface is really ontological and the region inside is just a representation of that surface created by the observer. But what if there is another observer inside the region, observing a smaller region through another bounding surface? Should we then say that the inner observer is not real? What about the case where two observers observe one another through different bounding surfaces - whose surface trumps whose? Moreover we can't simply put all possible bounding surfaces into the ontology, because then the ontology would contain the whole of spacetime and we would be back to the 3D picture. So evidently if we don't wish to resort to operationalism we will need some principled approach which allows us to identify an ontology of surfaces without appealing to external observers, in order that we can integrate observed systems and observers into a unified physical description.

There are two obvious ways to achieve this. The first is to accept that the choice of surface is always relativized to an observer, and adopt  a relational approach similar to relational quantum mechanics (RQM)\cite{sep-qm-relational,1996IJTP...35.1637R} in which we say that \emph{all} valid physical descriptions must be relative to an observer. RQM avoids collapsing to operationalism because it tells us that every physical system counts as an observer and thus each physical system defines its own observer-system split. So we could potentially add to RQM the specification that for every physical system, the description of the world relative to that system at a given time is defined on a two-dimensional surface, and that surface contains everything that belongs to the ontology defined relative to that observer\footnote{It is not entirely clear how one arrives at a specification of which observers exist in the first place within this kind of approach, given that everything is supposed to be relativized to an observer so there can't be observer-independent facts about the set of observers. But this is a generic problem for all relational and perspectival approaches to quantum mechanics - adding the specification that data is defined on surfaces doesn't make the problem worse, although it doesn't offer a solution either.}. Moreover, provided that we include a postulate ensuring the existence of links between the perspectives of different observers, as described in ref \cite{https://doi.org/10.48550/arxiv.2203.13342}, RQM still gives rise to a  shared macroscopic reality common to large classes of observers, and thus we can potentially arrive at something close to an integrated description of `the universe as a whole,' even though this approach does not allow such a thing as a completely observer-independent gods-eye view of the universe.

Alternatively, we could define a unique surface, or small set of surfaces, to which all of spacetime can be reduced - that surface  thus acting as something like a `hologram' (in Part II we will take a more detailed look at holography).  For example, a picture like this arises within  Kent's solution to the Lorentzian quantum reality problem\cite{Kent,2015KentL} which is intended as a solution to the measurement problem: in this model the wavefunction undergoes its usual unitary evolution until the end of time, and then we imagine something like a measurement being performed on the final state, with the actual course of history being determined by the result of the measurement (the `measurement' is to be understood as a mathematical device for extracting probabilities, rather than a literal physical operation)\footnote{Kent also makes allowance for the possibility that there is no end of time - in this case we simply take a limit as $t \rightarrow \infty$, making some assumptions which ensure that the limit is well-defined.}. This approach  looks like it could be a good fit with the covariant bound because it has  the consequence that the information in a region (e.g. on a light-sheet) depends on the state of its bounding surfaces in its lightlike or timelike future: a beable  can exist at some spacetime point only if there is a record of its existence somewhere in the final state, and thus the total amount of information in a given region cannot be greater than the amount of information which can escape that region to arrive at timelike infinity, so something like the covariant entropy bound will naturally arise\footnote{The version of this approach presented by Kent in ref \cite{2017Kent} is slightly different because it has the consequence that the information at a point is determined only by parts of the final state which are \emph{outside} the future light-cone of that point, so this formulation of Kent's view does not postulate the same tight connection between the information in a region and its bounding surfaces in the lightlike or timelike future. In future work we hope to consider in more detail the relationship between entropy bounds and these two formulations of Kent's approach.}. And in this picture we have no need to appeal to external observers to decide which surfaces should feature in the ontology, because in a sense the ontology includes only the surface on which the result of the final measurement is defined; the contents of the rest of spacetime, including observers themselves, are merely a projection from that surface.

 \section{Gravity \label{evidence}}

   Clearly if we had arrived at the covariant entropy bound purely by counting the number of degrees of freedom in the observations that we make about various physical systems, the thought-experiment in section \ref{thought} would be a strong argument against taking the bound too seriously - after all, if  you know in advance that regardless of how many degrees of freedom there are  in a system  you will never be able to extract more than $n$  bits of information about it, then when you subsequently observe that you extract $n$ bits of information, you have grounds to conclude only that the system has at least $n$ degrees of freedom, not that it has exactly $n$ degrees of freedom.  However, we did not actually arrive at the covariant entropy  bound by directly counting degrees of freedom, but rather by a variety of theoretical arguments. This seems quite puzzling - our purely theoretical arguments are not subject to epistemic restrictions, so why should they have led us to the exact same bound that our epistemic restrictions would have imposed if we had literally counted degrees of freedom?
   
  In the case of the original thermodynamic arguments for the bound   this puzzle is quite easily resolved: we have already noted that thermodynamic entropy is relativized to a choice of manipulable variables, so  it would not be surprising if  the thermodynamic arguments for the Bekinstein-Hawking entropy and strong entropy bound were to lead us to a general bound  on \emph{accessible} degrees of freedom in regions of spacetime. However, we now have evidence for the bound based not on thermodynamics but on gravitational physics - for in many cases it is phenomena like gravitational collapse and gravitational focussing which protect the covariant bound.   In Bousso's own words: `\emph{Gravitational backreaction plays a crucial role in preventing violations of this bound. In realistic systems, an increase in entropy is accompanied by an increase in energy. Energy focusses light rays by an amount proportional to G. Thus it hastens the termination of a light-sheet at caustic points, preventing it from “seeing” too much entropy.}'\cite{Bousso_2004}  This leaves us with something of a dilemma: as we saw in section \ref{thought} there is a sense in which the covariant bound looks like it could correspond to an epistemic restriction, and yet some of the evidence for it comes from gravitational effects, which are surely not epistemic. Does this disprove the epistemic view, or can we argue that some epistemic considerations are somehow being smuggled into the gravitational evidence?

  \subsection{Focussing}

To answer this question, let us take a closer look a the focussing theorem. The theorem arises from Raychauduri's equation\cite{2007Prama..69...15E}, which describes the evolution of a `null congruence' (i.e. a family of geodesics) with affine parameter $\lambda$ in a D-dimensional manifold: 

\[  \frac{d \theta}{d \lambda} = -\frac{1}{D - 2} \theta^2 - \sigma_{ab} \sigma^{ab} -  R_{ab} k^a k^b \]

Here, $\theta$ is the expansion of the bundle, representing the change in the area spanned by infinitesimally neighbouring geodesics:  $\theta = \lim_{A \rightarrow 0} \frac{1}{A} \frac{dA}{d\lambda}$. $\sigma_{ab}$ is the shear, representing the change in the shape of the area spanned by infinitesimally neighbouring geodesics. Using Raychauduri's equation, we can arrive at the focussing theorem\cite{PhysRevD.93.064044}: in a spacetime satisfying the null curvature condition ($R^{ab}k^a k^b \geq 0$ for all null vectors $k^a$) the expansion is non-increasing at all points where $\theta$ is finite (i.e. points which are not singularities or caustics) of a surface-orthogonal null congruence. That is to say, if the null curvature condition is satisfied geodesics can focus, i.e. come together to a point, but they can never anti-focus, i.e. diverge away from one another. 

The focussing theorem is a theorem of differential geometry, so it does not depend on the Einstein equations or on any features of general relativity. However, Einstein's equations imply that the null curvature condition is equivalent to the null energy condition, i.e. $  T_{ab}  k^a k^b \geq 0$. So in a general relativistic spacetime, we can conclude that if the null \emph{energy} condition is satisfied then geodesics can focus but not anti-focus. And since the null energy condition is true at least on average for all realistic classical matter, we can conclude that in classical general relativity geodesics will always focus. 

It is evident that the covariant bound would not work without gravitational focussing. In Bousso's own words, `\emph{Entropy costs energy, energy focusses light, focussing leads to the formation of caustics, and caustics prevent light-sheets from going on forever.}'\cite{Bousso_2002} This leads to a natural speculation - could we  turn this reasoning around and explain gravitational phenomena as a consequence of an entropy bound? For if we are prepared to take for granted that the information on a light-sheet (or at least, the \emph{accessible} information on a light-sheet) is upper bounded by the information that can pass through its bounding surface, then Bousso's reasoning can be employed to argue that geodesics must be focussed by the presence of mass-energy, so the existence of some gravity-like behaviour is an inevitable consequence of the covariant entropy bound.

One may  wonder exactly how far we can get with this approach:  does the covariant bound single out the Einstein equations specifically, or does it merely lead us to a class of theories all exhibiting some sort of focussing of geodesics? In fact there have been suggestions that entropy-area relations can be used to derive the Einstein equations - most famously  Jacobson's 1995 `thermodynamical' derivation of the Einstein equations\cite{Jacobson_1995}. Jacobson's argument was published four years before Bousso published his covariant bound, so his derivation associates an entropy with a surface rather than a light-sheet; however, in appendix \ref{proof} we suggest a way of rewriting Jacobson's derivation using a light-sheet formulation where we suppose that the covariant bound is the consequence of a restriction on the number of degrees of freedom on a light-sheet, The derivation proceeds by assuming that   the resulting inequality relating the curvature of spacetime and area is saturated, i.e. spacetime is always curved just enough to ensure that the entropy bound is obeyed, and no more. The derivation  is not fully general - in appendix \ref{limit} we discuss some of its limitations - but it provides at least some support for the idea that   Einstein's equations could be regarded as a consequence of the covariant entropy bound, interpreted either ontologically or epistemically.

In what follows we will not assume that the exact form of the Einstein equations can be derived from the covariant entropy bound, but we will take it that, as indicated by Bousso's comments above, the covariant bound does at least entail the existence of some kinds of gravitational phenomena. So we have a two-way relation:  the covariant bound implies the existence of certain gravitational phenomena, and meanwhile certain gravitational phenomena imply the covariant bound, or at least play a major role in protecting the bound in many cases. It seems natural, therefore, to suppose that one of these things can be regarded as explaining the other - but how do we decide in which direction the explanation should go? 

Ref \cite{Chirco_2014} addresses this question, criticizing Jacobson's proof on the grounds that we must invoke the Einstein equations to arrive at  the universal relation between entropy and area on which the proof is based, so the derivation is circular. The authors assume that the entropy  is an entanglement entropy and then note that the entanglement is usually thought to depend on the   types of fields present, via the relation $S = \frac{n A}{L^2}$, where $n$ is the number of types of fields and $L$ is the cutoff length at which we reach the Planck regime. But matter fields interact gravitationally and thus if there are more fields present the energy density will be higher, so we will reach the Planck scale at a higher cutoff length $L$; it can be shown that the two effects exactly cancel out, so we end up with a universal relation between area and entropy that does not depend on the number of fields. Evidently if this is the right way to think about the relation between entropy and area, then this relation is a \emph{consequence} of the Einstein equations, and thus it would indeed be circular to use it to explain gravitational effects.

In part II of this paper we will examine the relationship between entropy bounds and entanglement entropy in greater detail, but for now it is enough to note that this criticism is begging the question: someone who believes that the covariant bound should be understood as explaining the existence of gravitational phenomena would presumably argue that the entropy bounds should \emph{not} be understood as a consequence of the Einstein equations, but should be accepted as true for some other reason - for example, because they are regarded as a fundamental fact about  ontology. From this point of view  the reasoning of ref \cite{Chirco_2014}  is the wrong way round: it is precisely \emph{because} there is a universal relation between entropy and area that the gravitational interaction between matter fields must act in such a way as to cancel the dependence on the number of fields, and indeed this can be regarded as simply an alternative version of the argument that a universal entropy-area relation entails the existence of gravitational phenomena.  

In fact, explaining gravitational phenomena by appeal to an entropy-area relation rather than vice versa arguably leads to a better explanation. For if we start from the Einstein equations and give an explanation similar to the one in ref \cite{Chirco_2014}, we have to conclude that it just happens to turn out that gravitational effects manage to cancel the dependence on the number of species, in a way that one might regard as suspiciously coincidental. The coincidence disappears if we run the explanation in the other direction, with gravity required to work in such a way as to respect the universality of the entropy bound. Moreover,  if we start from the Einstein equations and give an explanation similar to the one in ref \cite{Chirco_2014}, we have to accept that a large range of physical mechanisms are responsible for protecting the entropy bound in different circumstances: `\emph{they differ according to the physical situation studied, and they can involve combinations of different effects more reminiscent of a conspiracy than of an elegant mechanism.}'\cite{Bousso_1999} This piecewise explanation seems quite unsatisfactory, whereas we can give a more unifying explanation if we see  the entropy bound as a fundamental feature of the ontology and derive all of these effects from it. Therefore we do not think it can be taken for granted that the correct direction of explanation is from Einstein's equations to the covariant bounds rather than vice versa.  Let us now consider what the epistemic and ontological views of the covariant bound have to say about the matter.

\subsection{Epistemic View \label{epistemic2}}

If we take an epistemic view of the covariant bound, we know that gravitational phenomena do not explain the covariant bound, because the bound is already fully explained by the facts about how much information observers can obtain out of various regions of spacetime. So, if we accept that there is an explanatory relation between the covariant bound and gravitational phenomena,  we can straightforwardly conclude that gravitational phenomena must be explained by the bound. 
And furthermore, since it is plausible that scientific explanations are usually transitive (as argued by Lange in ref \cite{Lange2018-LANTSA-9}), it would seem  to follow that gravitational phenomena  must ultimately be explained by the facts about how much information observers can obtain out of various regions of spacetime.  This  provides an answer to the question of how we could possibly have obtained a bound which is really just an epistemic restriction on the amount of information we can obtain out of various regions of spacetime, even though we arrived at this bound by consideration of gravitational effects rather than explicitly counting the number of states in various regions. For if we understand gravitational phenomena as a consequence of the same epistemic restrictions which give rise to the covariant bound itself, then reasoning based on gravitational phenomena can be expected to lead to the same results as we would have obtained had we explicitly counted states.

Now, it is likely that  explaining gravitational phenomena  as a consequence of an epistemic restriction in this way would  invite criticisms similar to Albert's emphatic reaction to the idea that the thermodynamic entropy could be epistemic:
`\emph{Can anybody seriously think that it is somehow \textbf{necessary} . . . that the
	particles that make up the material world must arrange themselves
	in accord with \textbf{what we know}, with \textbf{what we happen to have looked into?}
	Can anybody seriously think that our merely being ignorant of the
	exact microconditions of thermodynamic systems plays some part in
	\textbf{bringing it about}, in \textbf{making it the case}, that (say) \textbf{milk dissolves in coffee}?
	How could that \textbf{be}?}'\cite{Albert2000-ALBTAC} 
	
	However,  in response to this kind of criticism it is important to recall that, as noted in section \ref{uncertainty}, `epistemic' is not the same as `subjective.' If indeed it is the case that only a certain amount of information about a system can emerge through its bounding surface, that is an objective fact about reality: so if gravitational phenomena are to be understood as a consequence of limitations on the amount of information that can pass through a surface, then they are indeed an objective fact about a certain regime of reality. From this point of view gravitational focusing is in a sense an illusion, but the fact that all realistic observers are subject to that illusion is very much objective. 

One might also criticize this approach on the grounds that the argument we have given  leans too heavily on some questionable assumptions about explanation: in addition to assuming the transitivity of scientific explanation, we have assumed that  once we have a complete explanation of some physical phenomenon, we shouldn't posit another one. But many philosophers believe that explanation must be relativized to an audience or context of explanation\cite{van1980scientific,10.1093/oso/9780190652913.003.0004}, so really this assumption should say that we shouldn't posit another   explanation \emph{relative to the same audience or context of explanation}, and thus one might hope to avoid the conclusion by arguing that in one context of explanation the covariant bound is explained by epistemic limitations, and in another context of explanation the covariant bound is explained by gravitational phenomena, so there is no single context of explanation within which we can apply the transitivity of explanation to infer that  gravitational phenomena are explained by epistemic limitations.  

However, this response may be blocked by bypassing the transitivity argument and showing directly that  epistemic limitations on the information obtainable out of various regions of spacetime can give rise to `illusory' gravitational effects.  To do so, we will make an argument inspired by Susskind's ideas in ref \cite{doi:10.1063/1.531249}. Take a screen with area $A$, and   an object $O$ behind the screen with surface area $A'$; assume the surface postulate is correct, so the information on the surface of $O$ has an upper bound proportional to $A'$, and  the information passing through the screen has an upper bound proportional to $A$. Following Susskind, we will also assume the constant of proportionality is the same for both the surface and screen and that information on the surface of $O$ saturates the bound.  Now let light rays carry the information about the state of the surface of the object towards the screen. Suppose all the information in some area element $dA'$ is carried by a bundle of light rays $B$, which intersect the screen orthogonally and which have cross-sectional area $dA$ by the time they reach the screen.  Suppose that  gravitational focusing does not occur for some particular area element $dA'_x$ on the object, i.e. $\frac{d \theta}{d \lambda}$ is positive for the corresponding bundle $B_x$. Since $\theta$ is zero at the screen, this implies that $\theta$ is negative for $B_x$ between the object and the screen, and hence the corresponding area element $dA_x$ on the screen is smaller than $dA_x'$, so some of the information on the area element $dA'_x$ will not be able to pass through the screen. Thus we may perhaps make the case that the \emph{perceived} area of $dA_x'$ will be less than its real area, since the external observer can only judge the area of $dA_x'$ on the basis of the information that emerges about $dA_x'$ through the screen; so an external observer will always perceive the area of any area element $dA'$ on the surface of the object as being less than or equal to the corresponding area $dA$ on the screen, and thus it will always look as though gravitational focusing has occurred. This argument suggests a way  in which  gravitational focusing could indeed be somewhat illusory: it might be understood as a function of the fact that we must always perceive objects as having less surface area than the screens through which we observe them.

\subsection{Ontological View \label{ontological}} 

The question of whether gravitational phenomena are explained by the covariant bound or vice versa in an ontological picture will ultimately hinge on the nature of the ontological picture one adopts. But as a concrete example, suppose we adopt an ontological view   where the ontology is composed entirely of bounding surfaces. In that case, the covariant bound is explained by the fact that light-sheets are nothing more than an alternative encoding of the information on bounding surfaces, and then gravitational phenomena are explained by the ontology of light-sheets: geodesics must be focussed by matter because otherwise the light-sheets would contain more information than their bounding surfaces, which is impossible if the light-sheets are just alternative representations of bounding surfaces. 

  Smolin proposes something of this kind when he describes the significance of his own formulation of the entropy bound: `\emph{Its role is to constrain the quantum causal structure of a quantum spacetime in a way that connects the geometry of the surfaces on which measurements may be made with a measure of the information that those measurements may produce ... the notion of area is reduced fundamentally to a measure of the flow of quantum information}'\cite{Smolin2001TheSA}. It seems that Smolin is imagining something like the following:  spacetime should be understood as a construction  - presumably on the part of human observers, or subconscious information-processing going on in their brains -  whose geometry and dimensionality is determined by the nature of the information that the observers receive from the outside world, which need not itself be spatiotemporal in the ordinary sense.  So in this picture, it is indeed the case that light-sheets are just representations of the information on surfaces. Markopoulou and Smolin\cite{https://doi.org/10.48550/arxiv.hep-th/9910146} have constructed a model of this kind, based on information flowing in a network of `holographic screens'  which take the form of a causal set, with screens composed by a quadruple of events.

It should be noted that the derivation of the Einstein equations from the covariant bound relies on various approximations about entropy (particularly that it can be regarded as a fluid) and it is likely that these approximations do not hold at the most fundamental level. Moreover, the classical focussing theorem does not carry over precisely to  quantum mechanics, because it relies on the null energy condition, which can be violated by physically reasonable states in a quantum field theory. Thus if gravitational phenomena are really to be explained as a consequence of the covariant bound, understood in an ontological sense, it is likely  that gravity and  spacetime structure should be thought of as not fundamental but emergent - the fundamental description of reality pertains to the underlying degrees of freedom, and the curvature of spacetime emerges in the limit as the entropy-fluid approximation and the classical focussing theorem become valid, as a kind of higher-level description of the way in which these degrees of freedom depend on each other. The ontological picture thus falls in line with a long tradition of proposals about  `emergent gravity' and `emergent spacetime' within both physics\cite{Verlinde_2017,Verlinde_2011,Jacobson_1995} and philosophy\cite{Knox2013-KNOESG}.

 \section{Philosophical issues}
 
Having examined the empirical evidence for the different possible views of the covariant entropy bound, we will now consider if there are philosophical arguments which might support one view or another. Philosophical arguments are relevant here because  the  epistemic and ontological views are naturally associated with certain metaphysical pictures. In particular, an epistemic view of the bound follows naturally if we are committed to a view that might be described as `orthodox reductionism,' (i.e. the idea that spacetime is completely filled with autonomous degrees of freedom from which macroscopic phenomena emerge in some appropriate limit) because the assumption of autonomy at each spacetime point leads naturally to the idea that there may be information inside any given spacetime region which is unable to get out of the region. Conversely  an ontological view of the bound requires us to accept that  the content of  a spacetime region is somehow dependent on its bounding surfaces in the past and future, which is hard to reconcile with the orthodox reductionist picture - it seems to entail that  the degrees of freedom at different spacetime points are not generically autonomous, since they depend in some sense on higher-level phenomena pertaining to bounding surfaces. So it seems that an orthodox reductionist would have little option but to take an epistemic view. 

 Now, one might feel that there is something worrying about this dialectic. A metaphysical conviction  (i.e. orthodox reductionism) combined with certain empirical results (i.e. the evidence that spacetime is discretized and/or that information flux through surfaces is bounded) apparently compels us to postulate a realm of unobservable physics which we can never observe or learn anything definite about. It may seem worrying that we end up committed to a view so ontologically excessive - from a naturalistic point of view, one might argue that  rather than simply accepting the existence of  invisible, undetectable physics, we should consider abandoning the metaphysical convictions that led to this conclusion.  A historical comparison may help make the point: in the case of the ether hypothesis, it was the metaphysical conviction that light must be propagated in some substance which forced scientists of the day to postulate an invisible omnipresent substance which was undetectable in all experiments, and in hindsight it is easy to see that the right thing to do was to abandon the metaphysical conviction. 
 
 That said, some care must be taken here, for the   reason to get rid of the ether was not simply the fact that it was not directly detectable - we can have good theoretical reasons to believe in things that are not directly detectable. Rather, the ether was ultimately discarded because it served no theoretical function:  there was no  reason to believe in it other than the metaphysical conviction that light must be propagated in a substance, so this metaphysical conviction was forcing scientists of the day to make their theories more complex for no increase in explanatory or predictive power. This suggests that in order to decide whether or not we should accept the unobservable degrees of freedom postulated by an epistemic account of the bound, it is important to assess what theoretical function they might serve. Evidently they do not \emph{currently} serve any theoretical function, since our best current theories   obey the covariant bound and thus they don't include any unobservable degrees of freedom, but that leaves open the question of whether we could increase the explanatory or predictive power of our theories by taking these degrees of freedom into account. 

On the one hand, we saw in section \ref{uncertainty} that the existence of these degrees of freedom could potentially explain some features of our current physics.  On  the other hand, no \emph{specific} details about those degrees of freedom can possibly be explanatory, since ex hypothesi the world outside the relevant surfaces cannot depend on those details in any way. And one might question whether an entity really has theoretical power if only its existence and no specific features of its state or configuration are relevant to the predictions of the theory.    So arguably the \emph{literal} understanding of these degrees of freedom required by orthodox reductionism is indeed not serving any theoretical function - we would perhaps be better off thinking of these degrees of freedom as something like gauge, in that their existence is a necessary feature of our representation of the theory but their specific configuration is unimportant. That said, ultimately an answer to this question must await a better understanding of what a theory which explicitly includes unobservable degrees of freedom might look like - it's possible that the existence of specific states and configurations of these degrees of freedom might play some crucial theoretical function that is not immediately evident in advance of writing down the theory, and if that were the case then we would clearly be justified in being committed to their existence despite the fact that they are not directly detectable.

On the other hand,  abandoning reductionism comes with its own conceptual problems. In particular, when $A$ can be derived from $B$ but also $B$ can be derived from $A$, how do we know the correct direction of explanation? Orthodox reductionism provides a simple answer  -  the smaller always explains the larger and not vice versa - so if we are moving away from reductionism we will need alternative ways to settle these questions.   We saw something of this dialectic in the discussion of Jacobson's derivation in section \ref{ontological}: the argument of ref \cite{Chirco_2014} implicitly invokes the reductionist intuition that higher-level effects like the universality of the entropy-area relation must be given constructive explanations based on smaller-scale phenomena, such as the gravitational behaviour of fundamental fields, and yet a supporter of the view that entropy bounds explain gravitational phenomena would presumably want to insist that the gravitational behaviour of the fundamental fields is a consequence of the universal entropy bound, which would mean that a higher-level feature of reality actively constrains the behaviour of the smaller-scale entities.  If this kind of reasoning is to become accepted in science, we will need better criteria for judging the fitness of non-reductionist explanations and for deciding the correct direction of explanation in disputed cases.

 \section{Conclusion} 
 
So, assuming that the covariant bound is in some sense correct at least at the semiclassical level, is it epistemic or ontological? For those who are happy to take a purely operational view where we aim only to describe the degrees of freedom which are accessible to observers, this question may not be very important, but for those who hope to do cosmology, describe the universe as a whole or simply understand what is really going on, the issue must be resolved. 

We have seen that on the one hand, the thermodynamic arguments for the bound can easily be understood in an epistemic way, but on the other hand the arguments based on gravitational focussing don't seem particularly epistemic, and thus to maintain the epistemic interpretation it would seem that we would have to adopt quite a radically revisionary view which sees gravity itself as being in some sense epistemic.   This provides motivation for an ontological view, but on the other hand we have argued that an ontological view will likely also be quite revisionary in its account of the nature of spacetime -  in order to avoid being undermined by the possibility of an epistemic view, ontological approaches must show what is wrong with the assumptions underpinning the epistemic account, which will most likely involve proposing an alternative ontology which is a better fit to the nature of the data available to us, such as an ontology based on surfaces rather than autonomous degrees of freedom in a three-dimensional space.

The issues discussed in this article also have interesting consequences for philosophy of science more generally. For we have emphasized the way in which  the physics we arrive at is limited by features of our epistemic situation, and the `epistemic' interpretation of the covariant bound presents us with a vision of a world in which these limits are actually very severe, so significant portions of reality may be forever beyond our grasp. This poses some new challenges for scientific realism. First, how do we distinguish between parts of our theories which reflect features of reality and parts of our theories which are just consequences of our epistemic situation? And having done so, what kind of attitude should we take to parts of physics that we suspect are in fact consequences of our epistemic situation? For example, if we conclude that the entropy bound is just epistemic, it is certainly a part of reality and in that sense is compatible with scientific realism, and yet it would seem that  we should not adopt a naively realist attitude towards its proposed count of degrees of freedom, since that count is known to be objectively wrong. We hope to explore some of these issues in future work. 

  \section{Acknowledgements} 
 
 Thanks to Jacob Barandes, Niels Linneman, Ted Jacobson, and the UWO philosophy of physics reading group for very helpful comments on this article.  This publication was made possible through the support of the ID \#62312 grant from the John Templeton Foundation, as part of the project \href{https://www.templeton.org/grant/the-quantum-information-structure-of-spacetime-qiss-second-phase}{‘The Quantum Information Structure of Spacetime’ (QISS)}. The opinions expressed in this project/publication are those of the author(s) and  do not necessarily reflect the views of the John Templeton Foundation.

\appendix
\section{Deriving Einstein's Equations\label{proof} } 
	
		Here we make a  small modification to Jacobson's 1995 derivation  of the Einstein equations from a horizon entropy bound, in order to argue   that the covariant entropy bound can be used to derive Einstein's equations. Note that  this derivation ultimately requires us to assume that the inequality relating the curvature of spacetime and the area is saturated, i.e. spacetime is always curved just enough to ensure that the entropy bound is satisfied, and no more. This allows us to turn the covariant bound into an equality, so we can use basically the same mathematics as Jacobson's original proof, which assumed an entropy-area relation of the form $S = \frac{A}{4 \hbar G}$. Of course the derivation would not work without some such assumption, since we cannot derive an equality from an inequality unless we assume the inequality is saturated.

	\begin{proof} 
		
	At a given point in spacetime, consider a  2-surface $\Omega$ of area $A$ whose past directed null normal congruence $\{ k^a\}$ to one side has vanishing expansion and shear at the surface (as noted by Jacobson, it is always possible to choose such a surface provided it is sufficiently small). The past-directed null normal congruence has negative or zero expansion, and thus is a light-sheet $L$; we parametrize the rays $\{ k^a\}$ on $L$ with an affine parameter $\lambda$  which vanishes at $A$. Because $\Omega$ is small, we will take it that the Ricci tensor $R_{ab}$ and the stress-energy tensor $T_{ab}$ are approximately constant in the neighbourhood of $\Omega$. We will assume that  $D$, the number of degrees of freedom on $L$, is less than or equal to $\frac{A}{4\hbar G}$, thus explaining the fact that the entropy on $L$ is upper-bounded by $\frac{A}{4\hbar G}$. 
		
		Now let $\Omega'$ be the  infinitesimally smaller surface obtained from $\Omega$ by moving an affine parameter distance $d \lambda$ along each light-ray. $\Omega'$ has area $A'$ and bounds the   light-sheet $L'$ which is infinitesimally smaller than $L$. We will similarly assume that $D'$, the number of degrees of freedom on $L'$, is less than or equal to $\frac{A'}{4 \pi G}$. Then we will make a locality assumption to the effect that the curvature of spacetime in the region sandwiched by $\Omega'$ and $\Omega$ depends only on the contents of this region, and in particular, it is independent of the  number of degrees of freedom on $L'$. Thus the curvature of spacetime between $\Omega'$ and $\Omega$ must be large enough that the entropy bound is  satisfied even if the number of degrees of freedom on $L'$ takes its maximal value, $\frac{A'}{4 \pi G}$. Therefore the number of degrees of freedom between  $\Omega'$ and $\Omega$ must obey the bound $D - D' \leq \frac{ A - A'}{4 G \hbar}$.  
		
		Note that $D - D'$ acts as an upper bound for the entropy that could be assigned by any external observer to this region, so this expression can be recognised as a form of the generalized version of Bousso's bound specified in ref \cite{Flanagan_2000}:  $S" \leq \frac{ A - A'}{4 G \hbar}$, where we cut off the light-sheet associated with a  surface of area $A$ before it terminates, at a point where it has area $A'$, and $S"$ is the entropy on the cut-off light-sheet. This bound is also known to be true in many circumstances, although not universally - its validity depends on the assumption that entropy can be regarded as something like a locally-defined fluid, so it will not necessarily hold in scenarios where the nonlocal nature of the entropy becomes important - and of course, in such scenarios our locality assumption about the number of degrees of freedom would not be expected to hold, so this is exactly the domain in which the derivation we suggest here would no longer be reliable.
		
		Let $\theta(x)$ denote the expansion of the light-rays in $L$ at value $x$ of the affine parameter $\lambda$;  thus   $A - A' = - \int_{  \Omega'} \theta(d\lambda) d A d\lambda$.
		
		Now we may employ the Raychauduri equation. Having chosen $\Omega$ such that the shear on it is zero, we have that  $\frac{d \theta}{d \lambda} = -\frac{1}{2} \theta^2 - R_{ab} k^a k^b$, where $R_{ab}$ is the Ricci tensor and $k$ is a unit vector field which at every point in $L$ is tangent to the light-ray in $L$ passing through that point,  oriented from $\Omega$ towards $\Omega'$.   Since  $\theta$ is zero on the surface $\Omega$, we can work in the limit of small $\theta$, and then since we have assumed that the Ricci tensor is approximately constant over the region, we infer that everywhere on $\Omega'$ we have $\theta = - R_{ab} k^a k^b d\lambda$. 
		
		Hence  $A - A' = - \int_{\Omega'} - R_{ab} k^a k^b d^2\lambda   dA  =   A' R_{ab} k^a k^b d^2\lambda $. 
		
		Now we must  derive a lower bound on $D - D'$. We will do this by assuming that the number of degrees of freedom in this region at least as large as any entropy that could be assigned to the region by an external observer.   It is evident that the entropy assigned by an external observer should be connected to the stress-energy tensor in some way; roughly speaking, higher energy systems have more possible states and hence higher entropy. But writing down an explicit expression will require some approximations. Here we will follow Flanagan, Marolf and Wald\cite{Flanagan_2000} in assuming that the relevant entropy can be expressed as an entropy flux $s^a(x)$, so that the entropy  between $\Omega'$ and $\Omega$ is given by $S = \int_{x \in \Omega'} s^a(x) k^a dx d\lambda$. As argued by ref \cite{Flanagan_2000}, this is a good approximation in many regimes, though not all. Bousso writes: `\emph{Generally speaking, the notion of an entropy flux assumes that entropy can be treated as a kind of local fluid. This is often a good approximation, but it ignores the non-local character of entropy and does not hold at a fundamental level}'\cite{Bousso_2002}.

		Moreover, from thermodynamics we have that $dS = \frac{dQ}{T}$, so we will take it that $s^a(x)$ can be written in the form of a heat flux divided by a temperature. Since $D - D'$ is at least as large as the entropy assigned by any observer, let us consider some particular  observer of acceleration $\kappa$  whose Rindler horizon coincides approximately with $L$ in this region. This observer sees a temperature  given by the Unruh temperature $\frac{\hbar \kappa}{2 \pi}$, as well as an energy flux given by the  boost-energy current of matter, $ T_{ab} \chi^b$, where $\chi^b$  is the (approximate) Killing vector field generating Lorentz boosts in the Rindler frame, which will coincide with a suitably defined null vector on the light-sheet. As noted by refs \cite{https://doi.org/10.48550/arxiv.1602.01474,Jacobson_1995}, in any spacetime, around any event, there exists a class of (possible) local Rindler observers who will perceive this boost-energy current and Unruh temperature, so we can be sure that such a possible observer will indeed be well-defined in the relevant region.

		As explained in  refs \cite{Jacobson2003-JACHE, Jacobson_1995}, $\chi^b$ and $k^b$ can be related as follows. Since we are performing an infinitesimal calculation to first order only, we can work on a stationary background. Then we have $\lambda k^a = \lambda (\frac{d v}{d \lambda} ) \chi^a$, where $v$ is the Killing parameter. In general the relation between affine and Killing parameters on a Killing horizon is $\lambda = a e^{\kappa v} $, where $a$ and $b$ are arbitrary constants, and $b$ can be chosen to be zero since we are free to shift the affine parameter. So we find that $\lambda \frac{ dv}{d \lambda} = \frac{1}{\kappa}$ and hence we conclude that $\chi^a = \lambda \kappa k^a$; thus at $\Omega'$ we have that $\chi^a = d\lambda \kappa k^a$. 
		
		Thus we infer that the entropy assigned by this specific observer is $S = 2 \pi \int_{ \Omega'} \frac{   T_{ab} k^a k^b} {\hbar} dA d^2\lambda$.  Since we have assumed that the stress-energy tensor is approximately constant over the region, we conclude that $S = 2 \pi  A' \frac{   T_{ab} k^a k^b} {\hbar}  d^2\lambda$, and hence $D - D' \geq  2 \pi  A' \frac{   T_{ab} k^a k^b} {\hbar}  d^2\lambda$.
  
  Thus in order to have $D - D' \leq  \frac{ A - A'}{4 G \hbar}$, we require:
		
		\[ 2 \pi  A' \frac{   T_{ab} k^a k^b} {\hbar}  d^2\lambda \leq    A' \frac{R_{ab} k^a k^b }{ 4 G \hbar} d^2\lambda \]
		
		and thus $ 8 \pi G       T_{ab} k^a k^b     \leq      R_{ab} k^a k^b   $.
		
		Let us now assume that the the bound is in fact saturated - i.e. space is always focussed exactly enough to ensure that the number of degrees of freedom obeys the covariant entropy bound, and no more. Thus $ 8 \pi G    T_{ab}  k^a k^b    = R_{ab} k^a k^b$. As noted by Jacobson, this implies that $8 \pi G T_{ab} = R_{ab} + f g_{ab}$ for some $f$.  Local conservation of energy and momentum implies that $T_{ab}$ is divergence free and therefore, using the contracted Bianchi identity, that $f = -\frac{R}{2} + \Lambda $ for some constant $\Lambda$. So we get $8 \pi G T_{ab} = R_{ab} -\frac{R g_{ab}}{2} + \Lambda g_{ab}$ which is the Einstein equation. 
		
	\end{proof}

	\subsection{Limitations of the derivation \label{limit}}
	
	The derivation given above has a number of limitations.  In particular, although it is clear that entropy must be related to the stress-energy tensor in some way, there is no universally valid expression giving the relation between them, so the `entropy flux' expression used in this calculation will not be valid in all situations - for example, it may fail to hold when the non-local character of entropy is relevant, or when matter is not in thermodynamic equilibrium. One might also question whether the Unruh temperature used in the calculation of the entropy is the right temperature to use here - what if the matter in the region between $S$ and $S'$ is just ordinary classical matter with a normal thermodynamical temperature not associated with any acceleration? Perhaps the Unruh temperature should be thought as encoding something to do with the maximum amount of entropy which could possibly be present for a given value of $T_{uv}$ (since after all it includes Planck's constant) but the details are hazy. But nonetheless,  it seems reasonable to think that a relation of a similar form must hold in other scenarios, since it is clear that entropy should be linked to the stress-energy tensor, so it seems plausible to think the argument should generalize. (Note that Jacobson did subsequently generalize his argument to the case of non-equilibrium thermodynamics\cite{Eling_2006} so one might think that similar strategies will work here).

	Also, we assumed at the final stage that the bound  $ 8 \pi G   T_{ab} k^a k^b     \leq      R_{ab} k^a k^b   $ is in fact saturated. This assumption  is somewhat questionable because we justified our entropy calculation on the basis that the entropy observed by the Rindler observer  must be a lower bound on the true number of degrees of freedom in the region - we have not ruled out the possibility that some other observer could see more entropy in the region, and if that were the case the true number of degrees of freedom would have to be larger. However, note that the specific choice of acceleration $\kappa$ cancels out during the entropy calculation, so the same calculation will work for any accelerated observer in the same location; so if one assumes that non-accelerated observers necessarily see less entropy than accelerated observers, one may argue that the entropy observed by the Rindler observer also provides a suitable \emph{upper bound} in this situation. 
	
	In addition,   we have assumed that the entropy inside the region is given only by the entropy of matter; but in principle there should also be entropy associated with gravitational degrees of freedom. Flanagan et al make this point in ref \cite{Flanagan_2000}, but they then leave out gravitational contributions because it's unclear how to calculate them. However, one  might worry that once gravitational degrees of freedom are included, a derivation of this kind will potentially be subject to some kind of infinite regression, and thus leaving out gravitational degrees of freedom is a non-trivial approximation. In Part II of this paper we will look in more detail at the question of whether or not an entropy bound can still hold once gravitational degrees of freedom become nontrivial.

\end{document}